# Method for Determining the Similarity of Text Documents for the Kazakh language, Taking Into Account Synonyms: Extension to TF-IDF

Bakhyt Bakiyev
*School of Computer Science*
*University of Birmingham*
Birmingham, United Kingdom
b.bakiyev@bham.ac.uk

*Abstract*—The task of determining the similarity of text documents has received considerable attention in many areas such as Information Retrieval, Text Mining, Natural Language Processing (NLP) and Computational Linguistics. Transferring data to numeric vectors is a complex task where algorithms such as tokenization, stopword filtering, stemming, and weighting of terms are used. The term frequency – inverse document frequency (TF-IDF) is the most widely used term weighting method to facilitate the search for relevant documents. To improve the weighting of terms, a large number of TF-IDF extensions are made. In this paper, another extension of the TF-IDF method is proposed where synonyms are taken into account. The effectiveness of the method is confirmed by experiments on functions such as Cosine, Dice and Jaccard to measure the similarity of text documents for the Kazakh language.

*Keywords—TF-IDF; Modified TF-IDF; Similarity measure; Cosine measure; Dice measure; Jaccard measure.*

## I. Introduction

For various purposes, the challenge of identifying the similarity of text documents has gained great interest in numerous areas such as Information Retrieval, Text Mining, Natural Language Processing (NLP), and Computational Linguistics. Search engines utilise it to find similar documents in answer to user requests, which is one of the most prevalent applications. The weighting of terms technique plays a critical role in this. The term frequency–inverse document frequency (TF-IDF) method is the most commonly used term weighting method for making it easier to find relevant documents. Traditional TF-IDF, on the other hand, does not take into account semantic connections between terms, which could result in more relevant content. Of course, there are other modern algorithms for doing this, but we will concentrate on TF-IDF in this work. Many extensions to TF-IDF have been proposed to date and they are used for various purposes. In this paper, the proposed method uses a TF-IDF modification that takes into account the word synonyms. The effectiveness of the method is confirmed by experiments on functions such as Cosine, Dice and Jaccard to measure the similarity of text documents for the Kazakh language. Synonyms are usually discovered using a thesaurus. The most popular thesaurus of synonyms is WordNet [22, 23]. However, there is no compiled thesaurus for the Kazakh language. As a result, a small dictionary of roughly 1000 Kazakh synonymous words was created for the project. The problem considered in the work is that it is necessary to compare two text documents for similarity. Let's say, documents are called as follows: document Q is a query, D is a document. First step is to prepare all the words to determine the similarity. This is done as follows: all unnecessary short words (pronouns, numbers, symbols, etc.) are removed, and then their stems are extracted from the remaining words. Usually, this is done using the Porter or Snowball algorithms, but for our work, we used the Kazakh rule-based stemming algorithm [3]. In the next step, the stems of the words are converted into vectors using TF-IDF method and as a result a table with numbers is formed and the tables with numbers (vectors) are built, the cosine or other measure is used to calculate the similarity of the documents. It is necessary to find a dictionary of synonyms of the Kazakh language in electronic form and add it to the work. So that, when determining the similarity of documents in the Kazakh language, the program must take into account the synonyms of the words of the Kazakh language for those cases where the search words do not match, since there is a possibility that their synonyms may match. Thus, a higher accuracy of determining the similarity of text documents in the Kazakh language using word synonyms can be achieved. To attain the aforementioned purpose, the following tasks are carried out and described in this paper:

- Analyze existing methods where TF-IDF function uses synonyms to determine the similarity of text documents;
- Modify the method for determining the similarity for the Kazakh language to allow the use of synonyms;

- Evaluate the performance of the proposed method in comparison with the existing method;

## II. RELATED WORK

Kumar et al. [1] used an approach of weighing terms based on synonyms for biomedical purposes. They present a novel Synonyms-Depending Term weighting scheme (SBT) that modifies Inverse Document Frequency (IDF) based on any term's synonyms-based cluster. They used MeSH to generate a dynamic cluster of synonyms for terms found in biomedical text sources. The IDF collaborates with the SBT to determine the similarity of biological words. The replacement of terms with their synonyms is triggered initially.

Gulic et al. [2] searched for synonyms of words within the same document and replaced them with general terms. They developed a matcher that employs the TF/IDF measure in conjunction with synonym recognition. The synonyms are determined at the beginning, and WordNet was used to do so. Their algorithm uses three strategies. 1)Identifies and replaces all synonyms in a document with a single common phrase. 2)Identifies and substitutes all synonyms in all papers with a single common phrase. 3)In both ontologies' texts, find all synonyms and replace them with a single common term.

The difference between our work and the related articles is that our suggested technique turns a word into its synonym during the process rather than at the beginning, and it only does so when TF-IDF fails to detect it, and both the TF and IDF components interact during that process.

## III. METHODS

In this section, a brief introduction of the approach is provided. A flowchart of the groundwork is illustrated in figure 1.

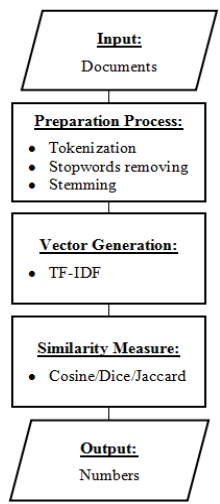

Figure 1. Flowchart of the groundwork

### A. Preparation Process

- Tokenization takes place at the word level. Punctuations and whitespaces are removed in this process except alphabetic characters or numbers. Numericals are relatively less relevant for the study and thus are removed.

- Stopwords removing is the process of removing common words like articles, prepositions, etc, like: a, an, the, in, on, under, off, out, with, etc. A list of 430 Kazakh stopwords for our work was built as shown in table 1.

TABLE I. LIST OF STOPWORDS IN KAZAKH.

| ай, айтпақшы, ал, алай, алайда, алақай, алатау, алдақашан, ана, анау, аһа, арбаң, …etc. |
|---|

- Stemming is the process of reducing a word to its root word. It is needed to have a dictionary of stemmed words. For this case, a stemming algorithm for Kazakh Language is adopted [3]. The Stemmer algorithm's performance was checked by comparing annotated Kazakh language corpus [4, 21] and as a result about 10000 Kazakh words have been stemmed for our work.

### B. Vector Generation

The well-known TF-IDF formulation has been used to compute weightings or scores for words.

TF-IDF weighting scheme [5, 6, 12-15, 17] is defined as:

$$(tf-idf)_{i,j} = tf_{i,j} \times idf_i$$

It assigns to term t a weight in document d that is:

- Highest when the term occurrence frequency is high within a small number of documents;
- Lower when the term occurrence frequency is fewer in a document, or occurs in many documents;
- Lowest when the term occurs virtually in all documents.

Term Frequency (TF), is defined as:

$$tf_{i,j} = \frac{n_{i,j}}{\sum_k n_{k,j}}$$

- where n is the number of occurrences of the considered term ($t_i$) in document $d_j$, and the denominator is the sum of number of occurrences of all terms in document $d_j$, that is, the size of the document $|d_j|$.

Inverse Document Frequency (IDF), is defined as follows:

$$idf_i = \log_2\left(\frac{|D|}{\{d:t_i \in d\}}\right),$$

- where D is the total number of documents, and in the denominator it is number of documents where the term $t_i$ appears. If the term is not in the document it will lead to a division-by-zero. It is therefore common to add one to the denominator.

### C. Similarity Measure

Many measures of similarity are specified nowadays, but the most known ones are Cosine, Jaccard, Dice measures [8, 18, 19, 20]. Although these measures are not shown to be the best, they demonstrate their value through many applications. Given two documents, a similarity function will examine how similar they are. The similarity function sim(vi,vj) is defined to compare two vectors vi and vj. This function should be symmetrical (namely sim(vi,vj)=sim(vj,vi)) and have a large value when vi and vj are somehow "similar" and constitute the largest value for identical vectors. A similarity function where the target range is [0, 1] is called a dichotomous similarity function. If the result is 0 then comparing two vectors vi and vj are not same. If the result is 1 it implies that they are exactly same. For the following and subsequent functions, [8] showed how to interpret the similarity measures between two vectors.

- Cosine measure or cosine coefficient, denoted as C is also symmetrical and is defined as:

$$\cos\theta = C(\vec{X},\vec{Y}) = C(\vec{Y},\vec{X}) = \frac{\vec{X}\bullet\vec{Y}}{|\vec{X}|\bullet|\vec{Y}|} = \frac{\sum_{i=1}^{n}x_i y_i}{\sqrt{\sum_{i=1}^{n}x_i^2}\sqrt{\sum_{i=1}^{n}y_i^2}}$$

- Jaccard measure denoted as J is also symmetrical and is defined as:

$$J(\vec{X},\vec{Y}) = \frac{\vec{X}\cdot\vec{Y}}{\|\vec{X}\|^2 + \|\vec{Y}\|^2 - \vec{X}\cdot\vec{Y}} = \frac{\sum_{i=1}^{n}x_i y_i}{\sum_{i=1}^{n}x_i^2 + \sum_{i=1}^{n}y_i^2 - \sum_{i=1}^{n}x_i y_i}$$

- Dice measure denoted as D is also symmetrical and is defined as:

$$D(\vec{X},\vec{Y}) = \frac{2(\vec{X}\cdot\vec{Y})}{\|\vec{X}\|^2 + \|\vec{Y}\|^2} = \frac{2\sum_{i=1}^{n}x_i y_i}{\sum_{i=1}^{n}x_i^2 + \sum_{i=1}^{n}y_i^2},$$

$x_i$ - components of vector X,
$y_i$ - components of vector Y,
n - length of vectors X and Y.

### D. Proposed Method

We propose a method which is an extension to the traditional TF-IDF to make the query extraction more effective which allows to get improved results over the traditional method. A flowchart of the proposed method is illustrated in figure 2.

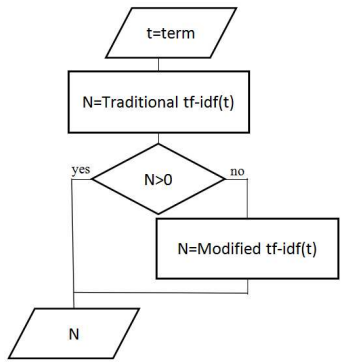

Figure 2. Flowchart of the proposed method

The algorithm deals with finding synonyms from a set of word synonyms stored in a two-dimensional array. Table 2 is an example of synonym words, where all synonyms of a word are recorded on one line; A0, B0, C0, and so on are words, and A0's synonyms are A1, A2, and so on; B0's synonyms are B1, B2, and so on. The algorithm detects the line where the synonym for the searched word is found, returns the synonym word, and then the Modified TF-IDF calculates N, if N is still equal to zero, the algorithm moves on to the next synonym word of the term. Therefore, it goes one by one into all synonym words in the line, from $0^{th}$ to last position synonym word until N>0, otherwise if no synonym word is found then N=0.

TABLE II. EXAMPLE OF TWO DIMENSIONAL ARRAY OF SYNONYMS.

|       |   | Columns |    |    |    |
|-------|---|---------|----|----|----|
|       |   | 0       | 1  | 2  | …  |
| Rows  | 0 | A0      | A1 | A2 | …  |
|       | 1 | B0      | B1 | B2 | …  |
|       | 2 | C0      | C1 | C2 | …  |
|       | … | …       | …  | …  | …  |

## IV. RESULTS

In order to compare similarity measuring functions (Cosine, Dice, Jaccard), the textual documents of 10 similar news about the blocking of telegram messenger (A1, A2, ...) and 10 similar news about oil prices in the world (B1, B2, ...) from the Internet in Kazakh are considered. Let's carry out the following documents using the traditional method. Following the application for the proposed method, the results are presented in figures and tables. In all figures, the prefix M stands for the word Modified, for example, M.Cos denotes the collaboration of Modified TF-IDF and Cosine measure, whereas Cos denotes the use of Traditional TF-IDF and Cosine measure. A score of 0.8, for example, indicates that document-1 is 80% similar to document-2, a score of 1 indicates that they are 100% similar, and a score of 0 indicates that they are completely dissimilar.

The results of the similarities A1 with A2-A10 documents (for similar documents) are given in graphical form in figure 3, where the overall results of the modified approach are higher than its counterpart.

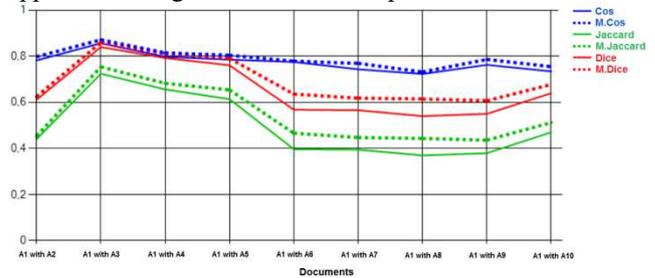

Figure 3. Results of the similar documents A1 with A2 - A10.

In addition, the results of the similarities B1 with B2-B10 documents (for similar documents) are graphically represented in figure 4. As shown in the previous figure, the modified method produces slightly better results than the traditional method in this figure as well.

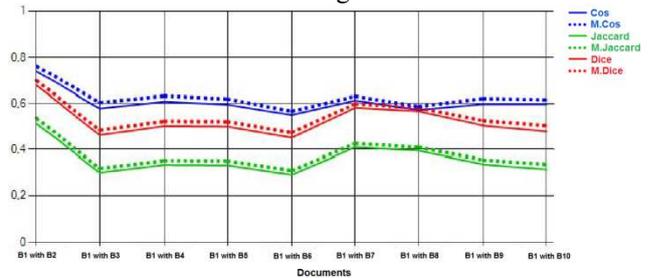

Figure 4. Results of the similar documents B1 with B2 - B10.

In figure 5, the results of the similarities A1 with B1-B10 documents (for dissimilar documents) are represented in graphical view. It is expected to yield a zero result for dissimilar texts because they are completely different documents; nonetheless, it is understandable that certain words may reappear, therefore results cannot always be zero. Also, you can see that the modified approach produces the same results as the traditional, which is a positive thing.

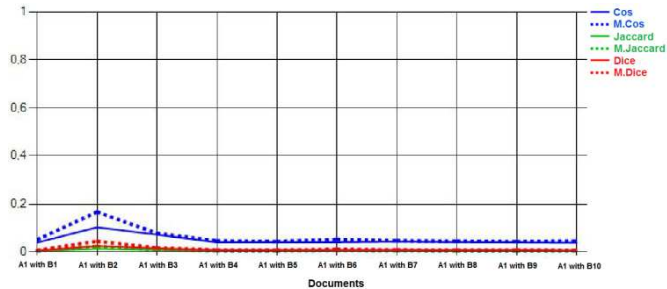

Figure 5.  Results of the dissimilar documents A1 with B1 - B10.

The results of the similarities B1 with A1-A10 documents (for dissimilar documents) are shown in graphical form in figure 6. As in the previous illustration, here also the overall outcomes of modified and traditional methods are very comparable, being close to zero.

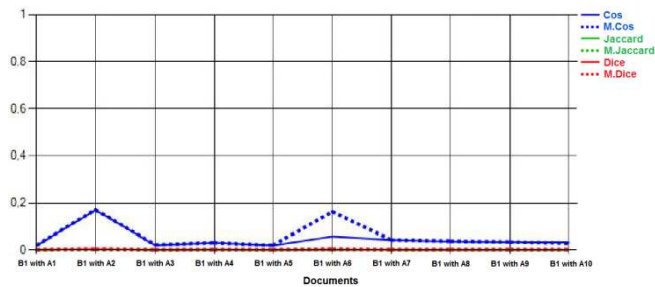

Figure 6.  Results of the dissimilar documents B1 with A1 - A10.

Next, table 3 shows the numerical representations of figures 3 and 4, whereas table 4 shows the numerical representations of figures 5 and 6. The average findings from tables 3 and 4 are combined in table 5 on the following page to show how the modified method outperforms its traditional counterpart in terms of overall efficiency. The difference between the averages produced from the traditional and modified approaches for similar documents in the blue background cells and for dissimilar documents in the green background cells is provided in table 5. Also, in that table "Modified TF-IDF minus Traditional TF-IDF" means that subtracting results of traditional TF-IDF from results of modified TF-IDF. In the cells with a yellow background, it is the difference between cells with a blue background and a green background. Thus, the modified method gives the values of similarity measures higher than its traditional counterpart.

TABLE III.  COMPARISON OF THE RESULTS OF SIMILARITY MEASURING FUNCTIONS FOR SIMILAR DOCUMENTS USING THE TRADITIONAL AND MODIFIED METHODS.

| Similarity measuring functions: | Cosine | | Jaccard | | Dice | |
|---|---|---|---|---|---|---|
| Documents | Traditional tf-idf | Modified tf-idf | Traditional tf-idf | Modified tf-idf | Traditional tf-idf | Modified tf-idf |
| A1 with A2 | 0,78 | 0,80 | 0,44 | 0,45 | 0,61 | 0,62 |
| A1 with A3 | 0,86 | 0,87 | 0,72 | 0,75 | 0,84 | 0,86 |
| A1 with A4 | 0,80 | 0,82 | 0,66 | 0,68 | 0,79 | 0,81 |
| A1 with A5 | 0,79 | 0,80 | 0,61 | 0,65 | 0,76 | 0,79 |
| A1 with A6 | 0,78 | 0,78 | 0,40 | 0,47 | 0,57 | 0,64 |
| A1 with A7 | 0,74 | 0,77 | 0,39 | 0,45 | 0,57 | 0,62 |
| A1 with A8 | 0,72 | 0,73 | 0,37 | 0,44 | 0,54 | 0,61 |
| A1 with A9 | 0,76 | 0,79 | 0,38 | 0,44 | 0,55 | 0,61 |
| A1 with A10 | 0,73 | 0,76 | 0,47 | 0,51 | 0,64 | 0,68 |
| **Average:** | **0,77** | **0,79** | **0,49** | **0,54** | **0,65** | **0,69** |
| B1 with B2 | 0,74 | 0,76 | 0,51 | 0,54 | 0,68 | 0,70 |
| B1 with B3 | 0,58 | 0,60 | 0,30 | 0,32 | 0,46 | 0,48 |
| B1 with B4 | 0,61 | 0,63 | 0,34 | 0,35 | 0,50 | 0,52 |
| B1 with B5 | 0,60 | 0,62 | 0,33 | 0,35 | 0,50 | 0,52 |
| B1 with B6 | 0,55 | 0,57 | 0,29 | 0,31 | 0,45 | 0,47 |
| B1 with B7 | 0,61 | 0,63 | 0,41 | 0,43 | 0,58 | 0,60 |
| B1 with B8 | 0,57 | 0,59 | 0,40 | 0,41 | 0,57 | 0,58 |
| B1 with B9 | 0,60 | 0,62 | 0,34 | 0,36 | 0,50 | 0,52 |
| B1 with B10 | 0,60 | 0,62 | 0,32 | 0,34 | 0,48 | 0,50 |
| **Average:** | **0,61** | **0,63** | **0,36** | **0,38** | **0,52** | **0,54** |

TABLE IV.  COMPARISON OF THE RESULTS OF SIMILARITY MEASURING FUNCTIONS FOR DISSIMILAR DOCUMENTS USING THE TRADITIONAL AND MODIFIED METHODS.

| Similarity measuring functions: | Cosine | | Jaccard | | Dice | |
|---|---|---|---|---|---|---|
| Documents | Traditional tf-idf | Modified tf-idf | Traditional tf-idf | Modified tf-idf | Traditional tf-idf | Modified tf-idf |
| A1 with B1 | 0,04 | 0,05 | 0 | 0 | 0 | 0,01 |
| A1 with B2 | 0,10 | 0,17 | 0,01 | 0,02 | 0,02 | 0,04 |
| A1 with B3 | 0,07 | 0,08 | 0,01 | 0,01 | 0,01 | 0,02 |
| A1 with B4 | 0,04 | 0,05 | 0 | 0 | 0,01 | 0,01 |
| A1 with B5 | 0,04 | 0,04 | 0 | 0 | 0,01 | 0,01 |
| A1 with B6 | 0,04 | 0,05 | 0 | 0 | 0,01 | 0,01 |
| A1 with B7 | 0,04 | 0,05 | 0 | 0 | 0,01 | 0,01 |
| A1 with B8 | 0,04 | 0,04 | 0 | 0 | 0,01 | 0,01 |
| A1 with B9 | 0,04 | 0,04 | 0 | 0 | 0,01 | 0,01 |
| A1 with B10 | 0,04 | 0,04 | 0 | 0 | 0 | 0,01 |
| **Average:** | **0,049** | **0,061** | **0,002** | **0,003** | **0,009** | **0,014** |
| B1 with A1 | 0,02 | 0,02 | 0 | 0 | 0 | 0 |
| B1 with A2 | 0,17 | 0,17 | 0 | 0 | 0 | 0 |
| B1 with A3 | 0,02 | 0,02 | 0 | 0 | 0 | 0 |
| B1 with A4 | 0,03 | 0,03 | 0 | 0 | 0 | 0 |
| B1 with A5 | 0,02 | 0,02 | 0 | 0 | 0 | 0 |
| B1 with A6 | 0,06 | 0,16 | 0 | 0 | 0 | 0 |
| B1 with A7 | 0,04 | 0,04 | 0 | 0 | 0 | 0 |
| B1 with A8 | 0,03 | 0,04 | 0 | 0 | 0 | 0 |
| B1 with A9 | 0,03 | 0,03 | 0 | 0 | 0 | 0 |
| B1 with A10 | 0,03 | 0,03 | 0 | 0 | 0 | 0 |
| **Average:** | **0,05** | **0,056** | **0** | **0** | **0** | **0** |

TABLE V. COMPARISON OF THE AVERAGE VALUE RESULTS OF SIMILARITY MEASURES OBTAINED USING THE TRADITIONAL AND MODIFIED METHODS.

| Row number | Documents | Cosine: Modified tf-idf minus Tradidional tf-idf | Jaccard: Modified tf-idf minus Tradidional tf-idf | Dice: Modified tf-idf minus Tradidional tf-idf |
|---|---|---|---|---|
| 1 | A1 with A2 - A10 (for similar documents) | 0,020 | 0,050 | 0,040 |
| 2 | A1 with B1 - B10 (for dissimilar documents) | 0,012 | 0,001 | 0,005 |
| 3 | 1st row - 2nd row: | 0,008 | 0,049 | 0,035 |
| 4 | B1 with B2 - B10 (for similar documents) | 0,020 | 0,02 | 0,02 |
| 5 | B1 with A1 - A10 (for dissimilar documents) | 0,011 | 0 | 0 |
| 6 | 4th row - 5th row: | 0,009 | 0,02 | 0,02 |

Finally, in table 5, all of the numbers in the cells with yellow, green, and blue backgrounds are positive, demonstrating that the proposed method works well. If these numbers were negative, it would indicate that the proposed strategy is ineffective. The values in the cells with yellow background gives us the information about how the proposed method works for similar and dissimilar documents. Positive values in the cells with yellow background indicates the proposed method for similar documents gives higher results than dissimilar documents which is another positive thing about the modified method.

## V. CONCLUSION

This paper proposes a method for determining the similarity of Kazakh text documents that takes synonyms into account, as an extension to TF-IDF. A comparison of the Cosine, Jaccard, and Dice similarity metrics revealed that the modified method outperforms the traditional method in all the above measures. The disadvantage was that the list of synonyms was short, because, as previously said, there was no electronic version of the list of synonyms in Kazakh. Even with our short collection of synonyms, the method performed admirably, and if we had a whole list of electronic synonyms, it would be even more effective.

As a future work, we plan to develop an algorithm that will generate a list of synonyms, and then we will compare this method to the related studies described before in this study [1, 2], as well as others, to further analyse the proposed method.

Our previously designated tasks, on the other hand, were completed, namely:
- The analysis of existing methods and algorithms where TF-IDF function uses synonyms for determining the similarity of text documents was carried out.
- A modification of the method for the Kazakh language was developed which uses synonyms as well.
- The performance of the modified method by comparison with its traditional counterpart was made.